\documentclass[a4paper,11pt]{article}
\usepackage{pos}

\title{Pion Distribution Amplitudes in the Continuum Limit}

\author*[a]{Nicholas Juliano}
\author[a,b]{Rui Zhang}
\author[c]{Carson Honkala}
\author[a,b]{Huey-Wen Lin}

\affiliation[a]{Department of Physics and Astronomy, Michigan State University,\\
        East Lansing, MI, 48824, U.S.A}

\affiliation[b]{Department of Computational Mathematics,
        Science and Engineering, Michigan State University,\\
        East Lansing, MI, 48824, U.S.A}

\affiliation[c]{Honors College, Michigan State University,\\
        East Lansing, MI, 48824, U.S.A}

\emailAdd{julianon@msu.edu}

\abstract{
We present a lattice-QCD calculation of the pion distribution amplitudes using large-momentum effective theory (LaMET). Our calculation is carried out using five ensembles with 2+1+1 flavors of highly improved staggered quarks (HISQ), generated by MILC collaboration, at 310 MeV and 220 MeV pion mass with 0.06, 0.09, 0.12 and 0.15 fm lattice spacings. We use clover fermion action for the valence quarks and tune the quark mass to match the lightest light and strange masses in the sea. The resulting lattice matrix elements are nonperturbatively renormalized in regularization-independent momentum-subtraction (RI/MOM) scheme and extrapolated to the continuum. We compare different approaches to extract the x-dependence of the pion distribution amplitudes.
}

\FullConference{%
 The 38th International Symposium on Lattice Field Theory, LATTICE2021
  26th-30th July, 2021
  Zoom/Gather@Massachusetts Institute of Technology
}

\begin{document}
\maketitle

\section{Introduction}

Meson distribution amplitudes (DAs) give the probability of finding a meson in a quark-antiquark Fock state. They are formally defined as follows, where $M(P)$ is some meson operator:
\begin{equation}
    \phi_M(x,\mu) = \frac{i}{f_M} \int \frac{d\mathcal{E}}{2\pi} e^{i(x-1)P\cdot\mathcal{E}n} \left< M(P)|\bar{\psi}(0)n \cdot \gamma \gamma_5 U(0,\mathcal{E}n) \psi(\mathcal{E}n)|0 \right>.
\end{equation}
Meson DAs are important for understanding how light-quark hadron masses emerge from QCD. They are also important inputs in many hard exclusive processes at large momentum transfers. In these processes, the cross-section can be factorized into a short-distance hard-scattering part and long-distance universal quantities such as lightcone DAs. The lightcone DAs can be determined from fits to experimental data or calculated from lattice QCD.

Meson DAs are like parton distribution functions in that they are universal quantities (meaning they are experiment and scale independent) and they are non-perturbative in nature. However, given meson composition compared to heavier hadrons, meson DAs are less constrained by experiments. They are also subject to various model-dependent calculations and we have no global-fitting result to compare them with.

\section{Lattice Setup}

Our lattices come from the MILC collaboration~\cite{Bazavov:2012xda}. We use 2+1+1 flavors of HISQs for the sea quarks, and clover fermion action for the valence quarks. On each lattice ensemble, we use multiple sources uniformly distributed in the time direction and randomly distributed in the spatial directions.

\begin{table*}[th!]
 \vspace{-2mm}
\center  
\begin{tabular}{|c|ccc|ccc|}
\hline
Ensemble ID    & $a$ (fm) & $M_\pi$ (MeV)    & $M_\pi L$   & $P_z$ (GeV)   & $N_\text{conf}$ & $N_\text{meas}$ \\\hline
$a15m310$    & 0.1510(20) & 320(5)      & 3.93      &  \{1.02, 1.54, 2.05\} & 452 & 10,848   \\
\hline
$a12m310$    & 0.1207(11) & 310(3)      & 4.55     & \{1.28, 1.71, 2.14\} & 1013 & 194,496   \\
$a12m220$    & 0.1184(10) & 228(2)      & 4.38     & \{1.31, 1.63, 1.96\} &  959 & 368,256   \\
\hline
$a09m310$    & 0.0888(8) &  313(3)      & 4.51     & \{1.31, 1.74, 2.18\} &  889 & 39,648   \\
\hline
$a06m310$    & 0.0582(4) &  320(2)      & 3.90      & \{1.33, 1.77, 2.22\} & 593 & 2,372   \\
\hline
\end{tabular}
\vspace{-2mm}
\caption{Information for different lattice ensembles used in this work. 
\label{tab:hisq}
  }
\end{table*}

Table \ref{tab:hisq} shows the five ensembles that we have done calculations on. We have four lattice spacings at 0.06, 0.09, 0.12, and 0.15 fm. We set the pion mass to 310 MeV at each lattice spacing, and additionally to 220 MeV at the 0.12 fm spacing. The overall momentum range is roughly 1.02-2.22 GeV. Also note the number of measurements that we have for each ensemble; for the lightest mass (the 220 MeV), we have the largest number of measurements (nearly 370,000).

To extract matrix elements, we first need to calculate the DA two-point correlators, which are defined for different mesons as follows:
\begin{equation}
    C_M^{DA}(z,P,t) = \left<0 \left| \int d^3 y e^{i \vec{P}\cdot \vec{y}} \bar{\psi}_1(\vec{y},t) \gamma_z \gamma_5 U(\vec{y},\vec{y}+z\hat{z})\psi_2(\vec{y}+z\hat{z},t) \bar{\psi}_2 (0,0) \gamma_5 \psi_1 (0,0) \right|0\right>
\end{equation}

\begin{figure*}
\centering
    \includegraphics[width=0.7\textwidth]{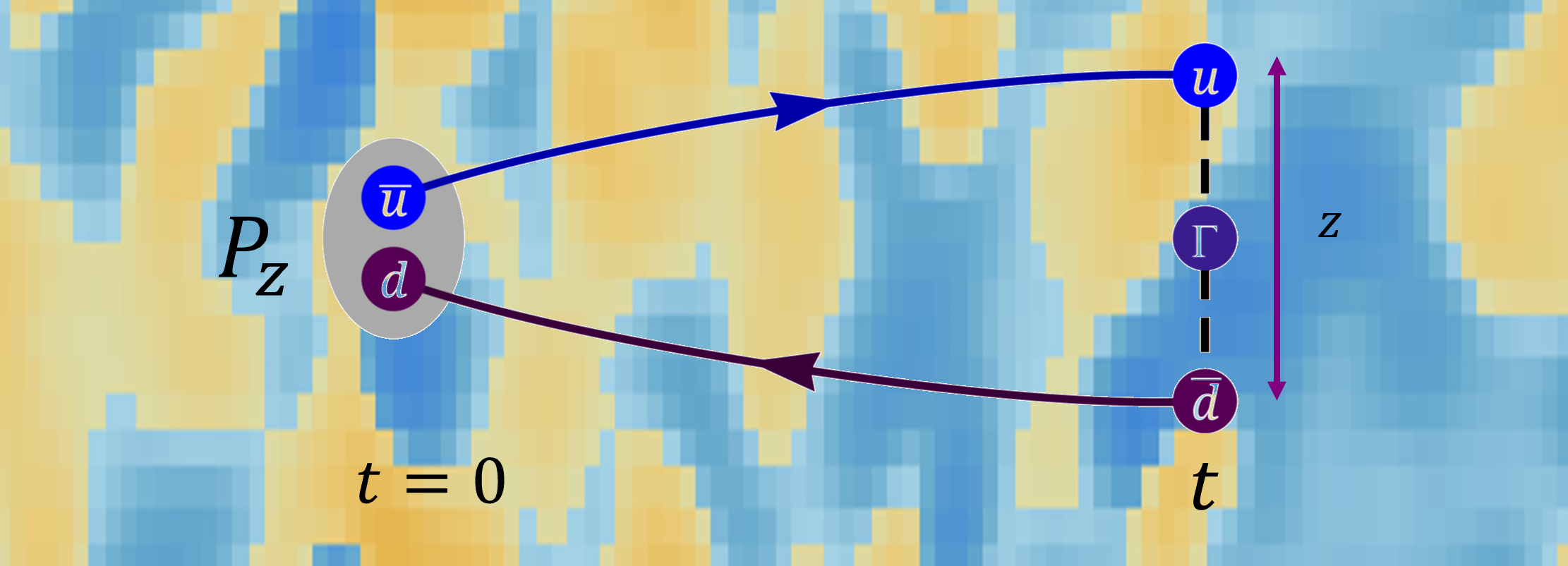}
    \caption{Diagram of our physical setup, illustrating how we construct our two-point correlators. In this work, we seek pion lightcone DAs.
    \label{fig:kaon_diagram}}
\end{figure*}

Figure \ref{fig:kaon_diagram} shows an example illustration of our physical setup, which gives context to our two-point correlators. `$\bar{\psi}_2 (0,0) \gamma_5 \psi_1 (0,0)$' is our local meson operator which creates the meson (a pion in this work).
`$\bar{\psi}_1(\vec{y},t) \gamma_z \gamma_5 U(\vec{y},\vec{y}+z\hat{z})\psi_2(\vec{y}+z\hat{z},t)$' is our main operator, which corresponds to the linked quarks.
Calculating these two-point correlators is the first step in obtaining our matrix elements, which are formally defined:
\begin{equation}
    \tilde{h}_M (z,P_z) = \left< M(P) \left| \bar{\psi}(0) \gamma^z \gamma_5 U(0,z) \psi(z) \right|0\right>
\end{equation}

\section{Matrix Elements}

\begin{figure*}
    \includegraphics[width=0.49\textwidth]{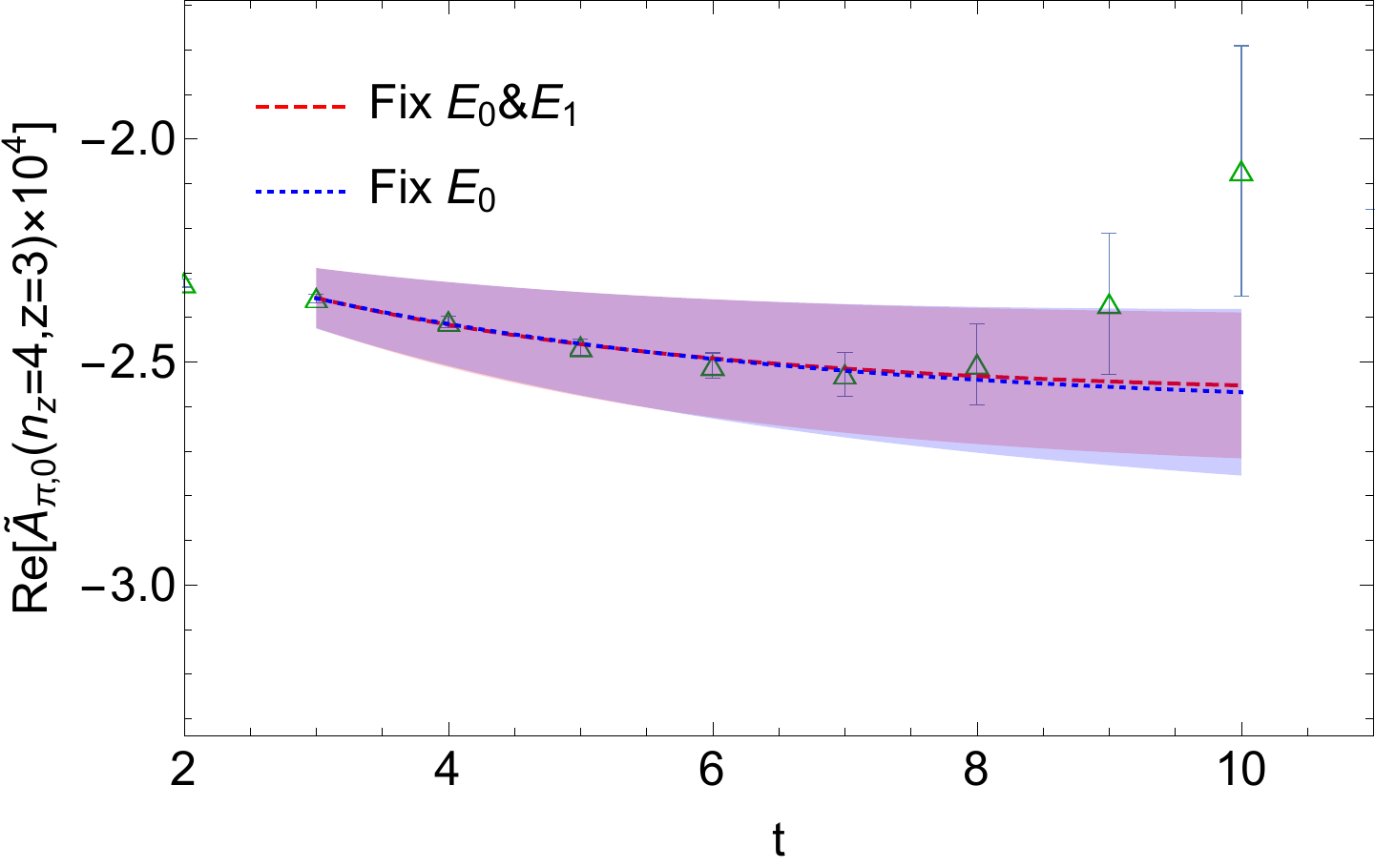}
    \includegraphics[width=0.49\textwidth]{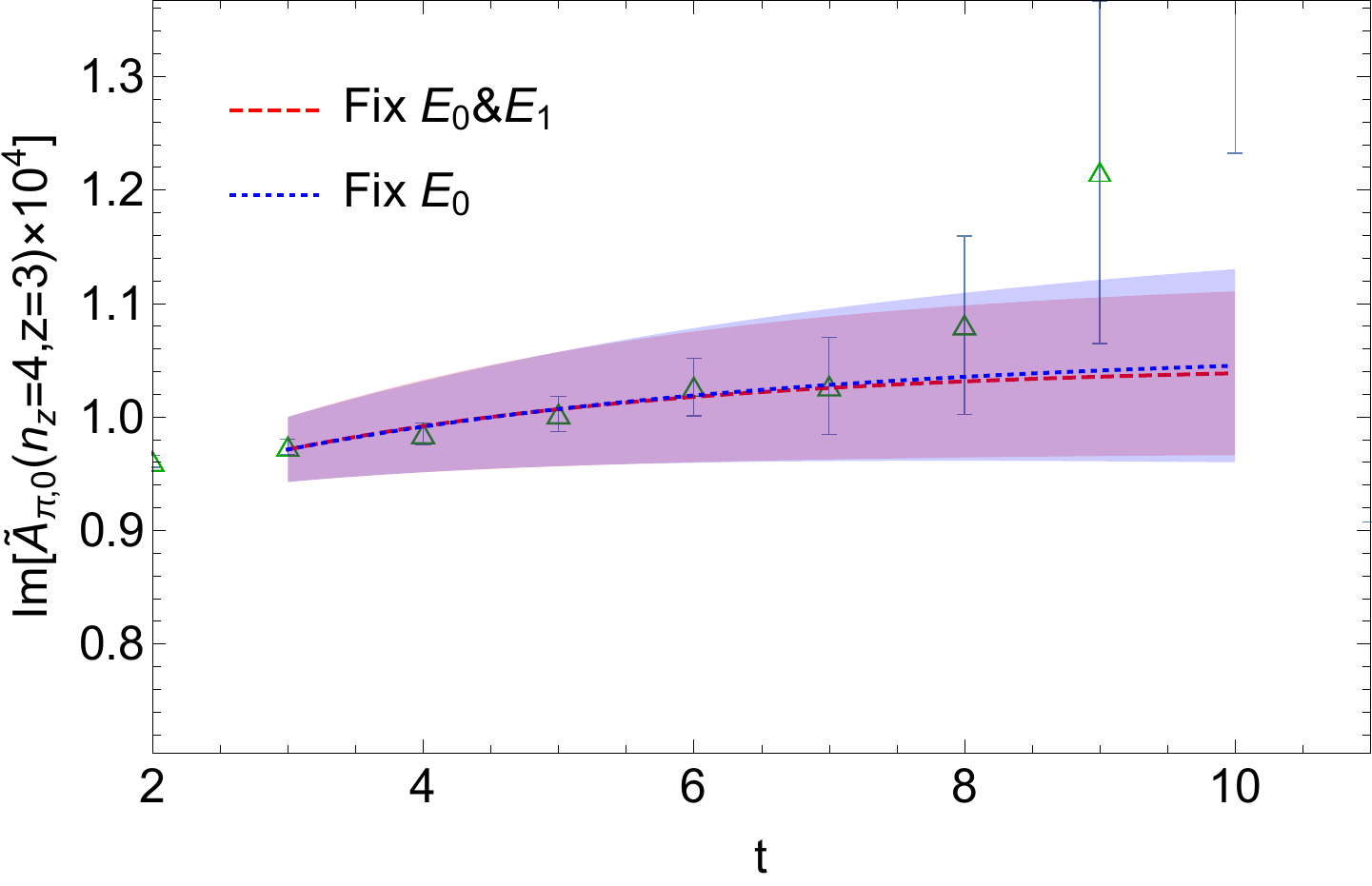}
    \caption{Two-point correlator fit test plots.
    \label{fig:correlator_fit}}
\end{figure*}

In practice, we extract the DA matrix elements from a two-point correlator fit to the following form:
\begin{equation}
    C_M^{DA}(z,P,t) = A_{M,0}^{DA}(P,z)e^{-E_{M,0}(P)t} + A_{M,1}^{DA}(P,z)e^{-E_{M,1}(P)t}+...
    \label{eq:correlator_fit_form}
\end{equation}
The amplitudes $A_M^{DA}$ of each exponential are proportional to the matrix elements $h$. Figure \ref{fig:correlator_fit} shows plots of $\tilde{A}$ versus time, where $\tilde{A}$ is the $C_M^{DA}$ correlator form in Equation \ref{eq:correlator_fit_form} multiplied by $e^{E_{M,0}(P)t}$ (the inverse of the ground state energy exponential) to minimize the overall $t$ dependence. The correlator data times that same exponential are plotted in green.

We use two fitting approaches to verify the stability of our results. First, we fix both the ground state and first excited state energies and fit the correlator; then, we separately fix only the ground state energy and fit the correlator. From the overlapping bands in the plots of Figure \ref{fig:correlator_fit}, we see a high level of agreement between both approaches. This tells us that our fit is stable.

\begin{figure*}
    \includegraphics[width=0.49\textwidth]{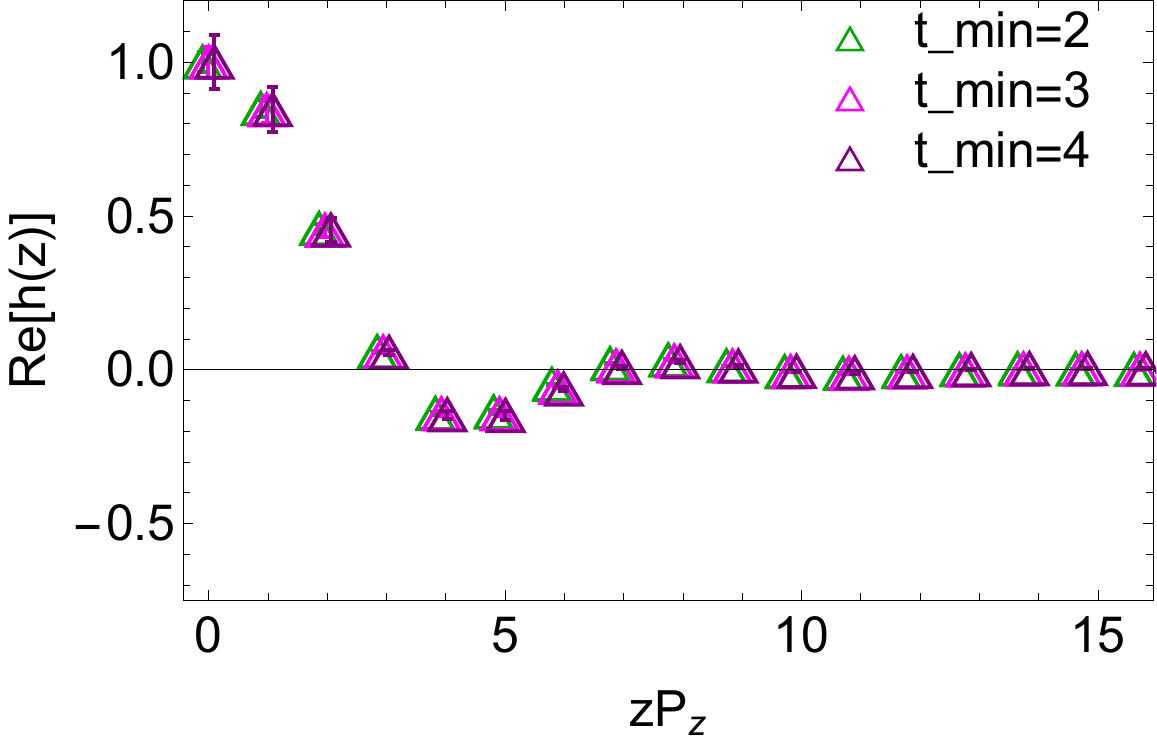}
    \includegraphics[width=0.49\textwidth]{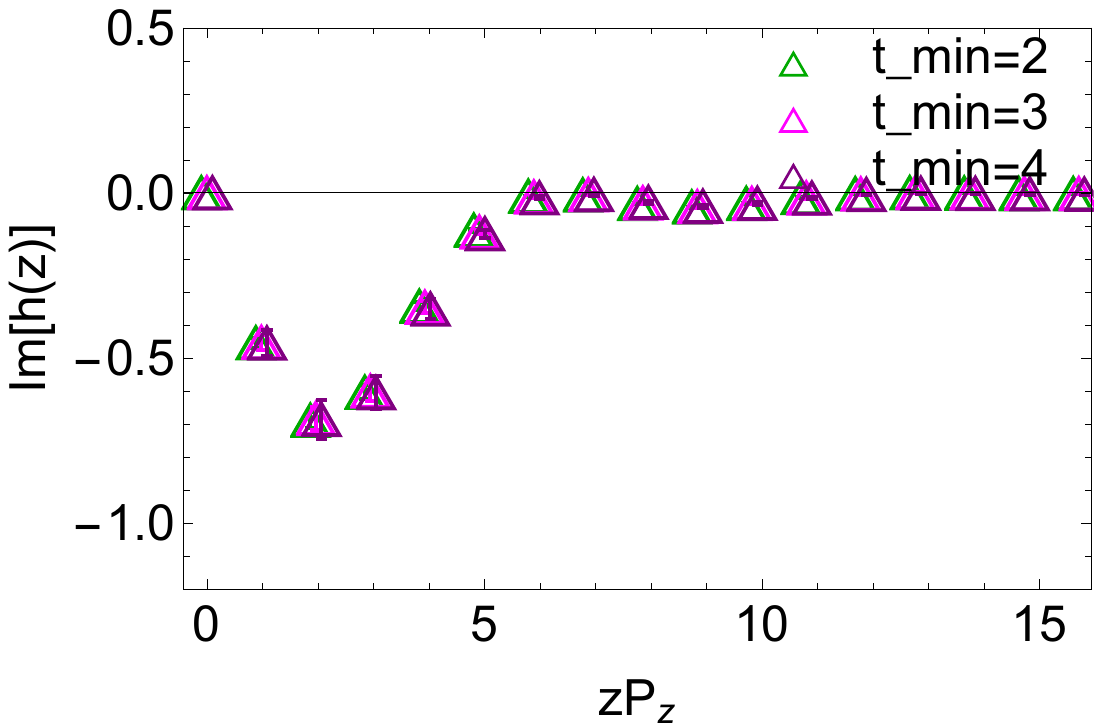}
    \caption{Real and imaginary bare matrix element plots for the a09m310 ensemble at $P_z$=2.18 GeV. We show three different fits with the minimum $t$ value varied between 2,3, and 4. The three sets show strong agreement.
    \label{fig:bare_me}}
\end{figure*}

As an example of what our fitted matrix elements look like, Figure \ref{fig:bare_me} shows $zP_z$ plots (‘$zP_z$’ being displacement times boosted momentum) of real and imaginary bare matrix element plots on the a09m310 ensemble at a momentum of 2.18 GeV. We vary the minimum $t$ value between 2,3, and 4 to determine if the resulting matrix elements show a strong dependence on $t_{min}$. In this case, even at a relatively high momentum, we see that the three sets of matrix elements all agree with each other, and in fact show very little difference from one another. This is another indication that our fit is stable.

To renormalize these matrix elements, we use the RI/MOM scheme. The details of the renormalization process are found in this group's prior work \cite{Zhang:2020}. We use the common scale $\mu^R$=3.8 GeV and $p_z^R$=0.

\begin{figure*}
    \includegraphics[width=0.49\textwidth]{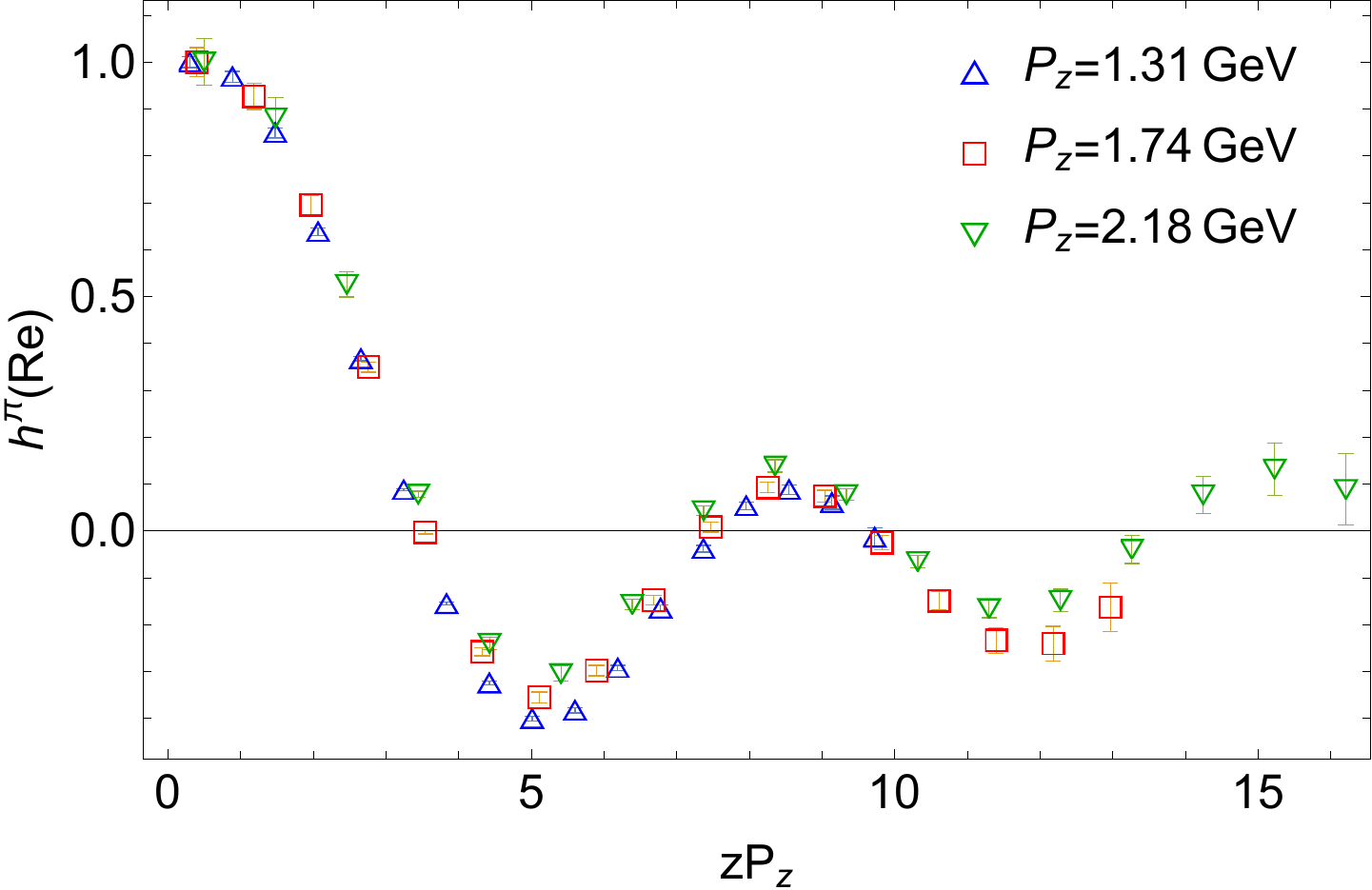}
    \includegraphics[width=0.49\textwidth]{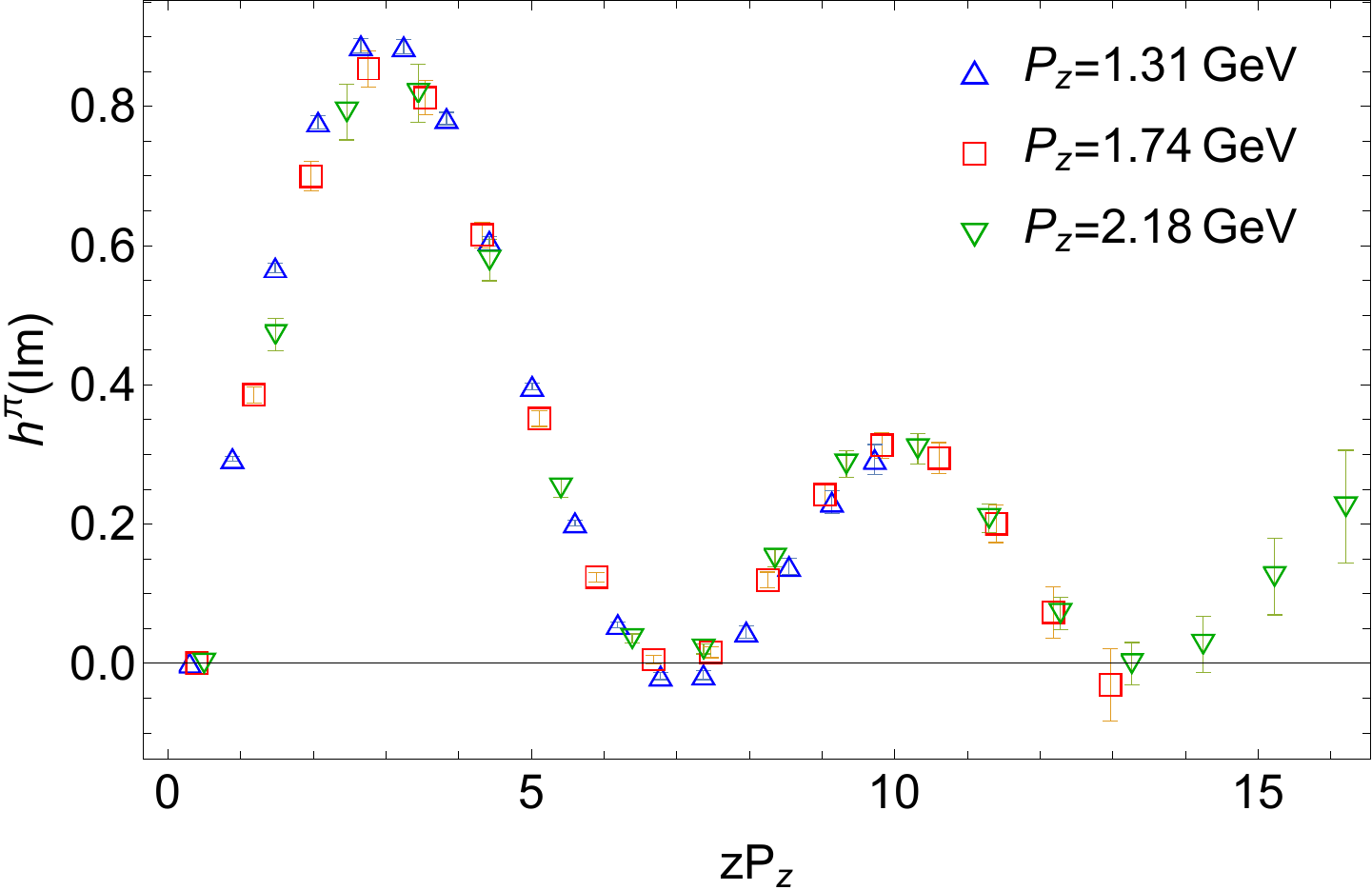}
    \caption{Real and imaginary $zP_z$ plots of our renormalized matrix elements on the a09m310 ensemble with $P_z$={1.31, 1.74, 2.18} GeV.
    \label{fig:renormalized_me}}
\end{figure*}

Still looking at the a09m310 ensemble (i.e., the lattice spacing and mass are fixed, isolating the momentum dependence), Figure \ref{fig:renormalized_me} shows real and imaginary $zP_z$ plots of our renormalized matrix elements. We choose three different values of $P_z$: 1.31, 1.74, and 2.18 GeV. There are small differences between the matrix elements at the three momenta, but overall, they reflect the same form. So we don’t see a strong $P_z$ dependence in our renormalized matrix elements.

\begin{figure*}
    \includegraphics[width=0.49\textwidth]{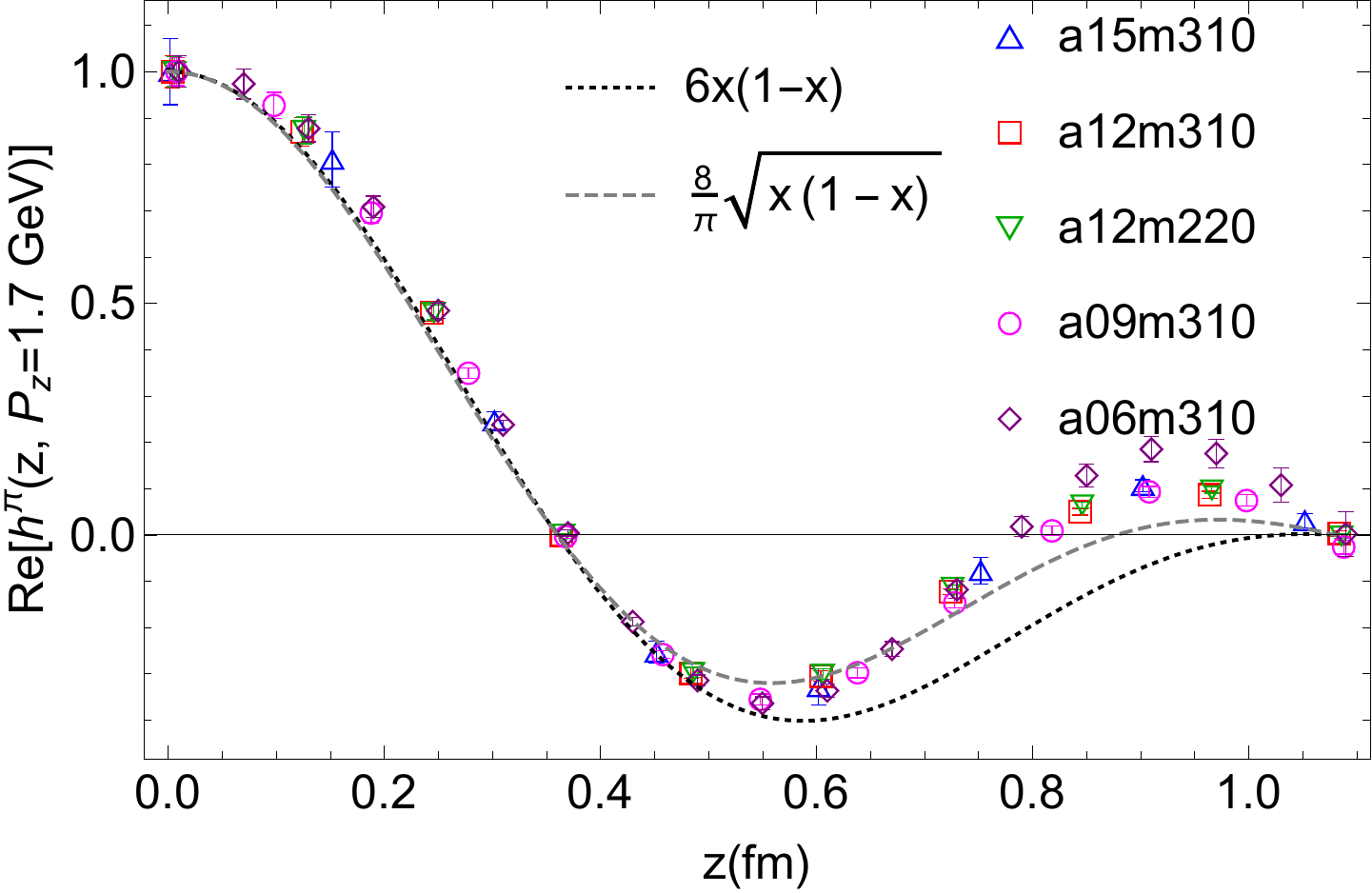}
    \includegraphics[width=0.49\textwidth]{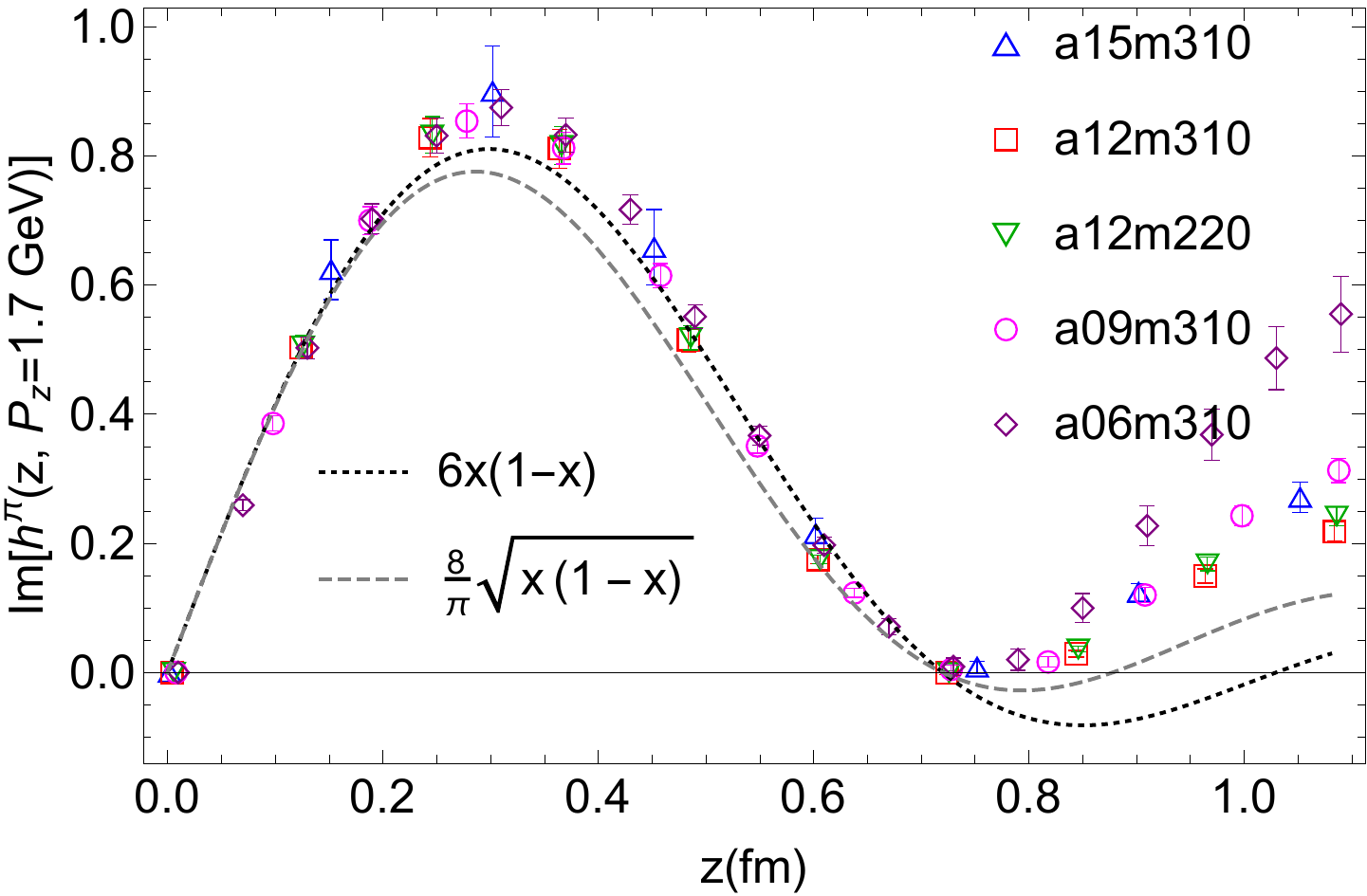}
    \caption{Real and imaginary $z$ plots of our renormalized matrix elements for all five of our ensembles with $P_z\approx$1.7 GeV.
    \label{fig:renorm_fixed_mom}}
\end{figure*}

Conversely, we isolate the mass and lattice spacing dependences by incorporating all five of our ensembles (see Figure \ref{fig:renorm_fixed_mom}). In this case we fix the momentum to be as close to 1.7 GeV as we can on each ensemble. The mass dependence comes from comparing the two fits at the 0.12 fm spacing: the green markers in Figure \ref{fig:renorm_fixed_mom} show the 220 MeV pion mass and the red markers show the 310 MeV pion mass. Both of these sets of matrix elements are very close to each other, indicating only minimal pion mass dependence from our fits.

For the lattice spacing dependence, we look at the purple (0.06 fm), pink (0.09 fm), red (0.12 fm) and blue (0.15 fm) markers in Figure \ref{fig:renorm_fixed_mom}. These four sets of matrix elements show general agreement in the small-to-mid-$z$ region, though we do see higher deviation than we did when looking at the mass dependence. Note that the disagreement we see in the large-$z$ region is likely due to higher twist effects, and so these results meet our expectations. The dashed and dotted curves are the Fourier transformations of two asymptotic approximation forms for the pion lightcone DA. We see that the real parts of the lattice data are closer to the $(8/\pi) \sqrt{x(1-x)}$ form, while in the small-to-mid-$z$ region, the imaginary parts are closer to the $6x(1-x)$ form.

\section{Pion Lightcone DA}

\begin{figure*}
    \includegraphics[width=0.49\textwidth]{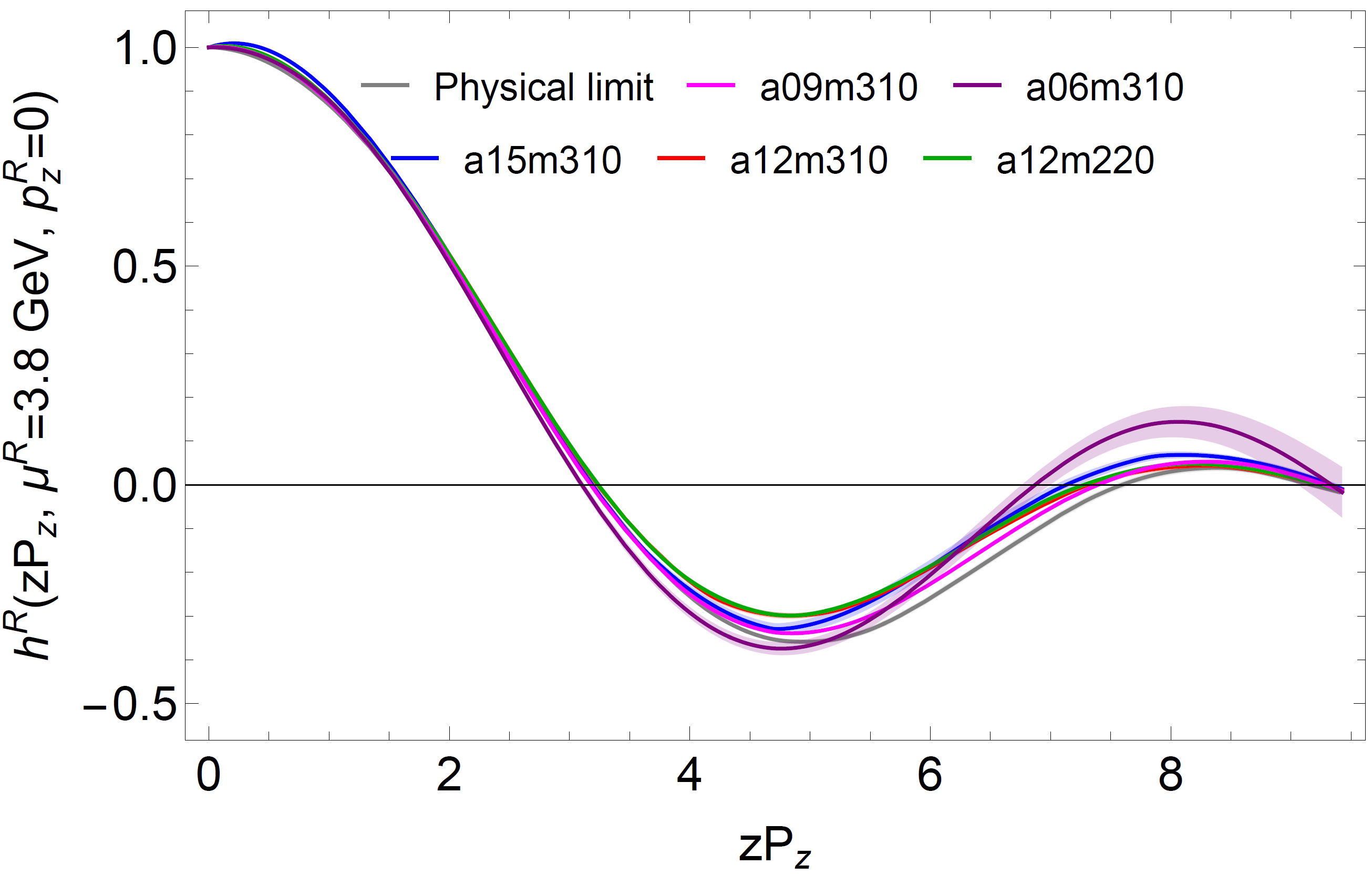}
    \includegraphics[width=0.49\textwidth]{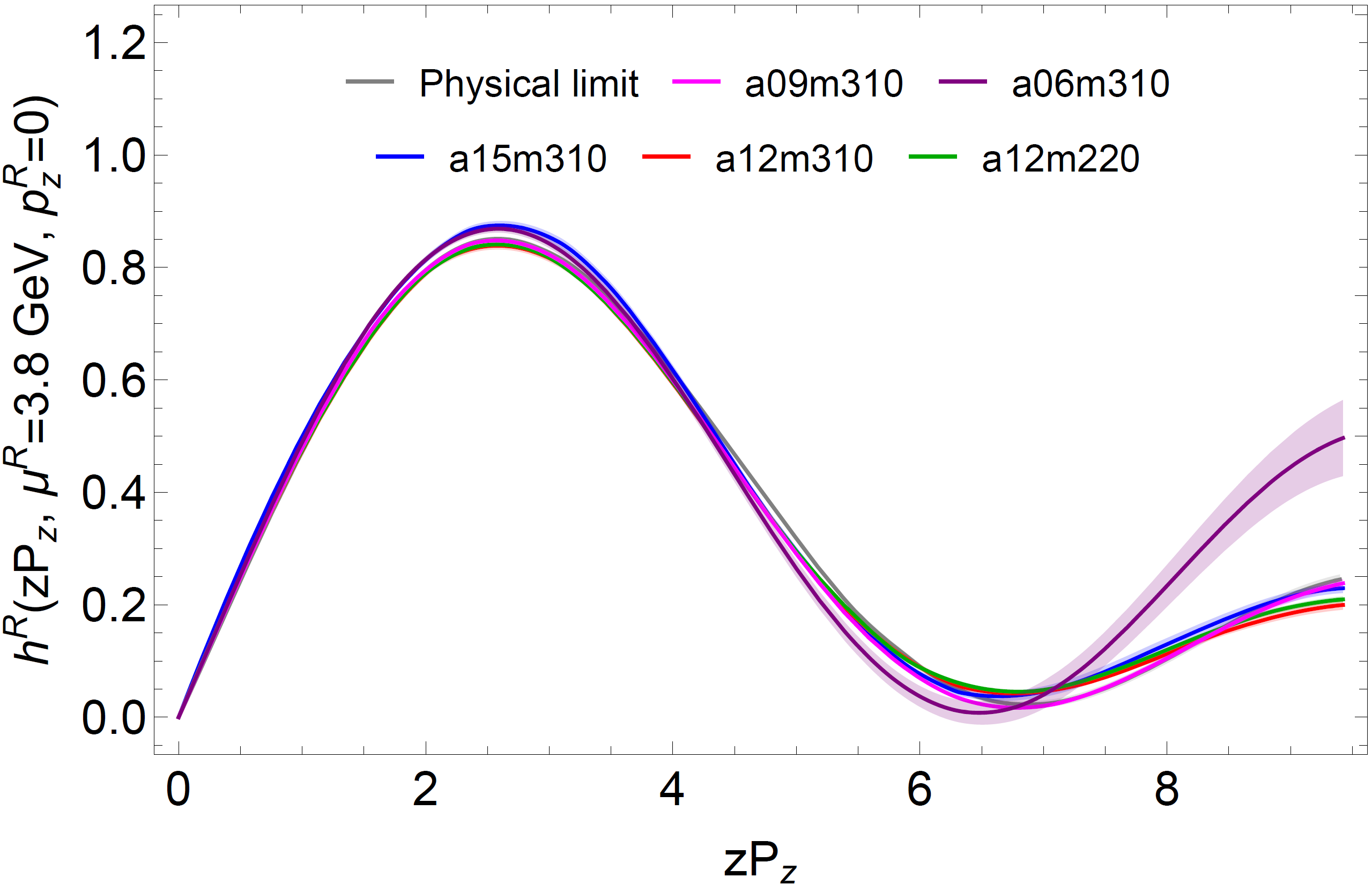}
    \caption{Real and imaginary plots of the renormalized matrix elements on all five of our ensembles with $P_z$=1.74 GeV. The physical limit is plotted in gray.
    \label{fig:renorm_phys_lim}}
\end{figure*}

Next, we look to determine the pion DA from the continuum-physical matrix elements. We use the following form to obtain these continuum-physical matrix elements $h$ through extrapolation:
\begin{equation}
    h = h_0 (1 + c_2 a^2 + d_2 M_{\pi}^2).
    \label{eq:extrapolation_form}
\end{equation}
Note that we use an $a^2$ lattice spacing dependence in this form. In the continuum-physical limit, our matrix elements become inputs into the following form: 
\begin{equation}
    h(z,\mu^R,p_z^R,P_z) = \int_{-\infty}^{\infty} dx \int_0^1 dy \;\; C \left(x,y,\left(\frac{\mu^R}{p_z^R} \right)^2,\frac{P_z}{\mu^R},\frac{P_z}{p_z^R} \right) f_{m,n}(y)e^{i(1-x)zP_z}
    \label{eq:continuum_physical_me}
\end{equation}
where $C$ is the appropriate matching kernel, which is calculated perturbatively \cite{Liu:2019urm}. Our goal is to determine the pion lightcone DA, which is some unknown function $f$ of momentum $y$ that we characterize by two parameters, $m$ and $n$. Figure \ref{fig:renorm_phys_lim} again shows $zP_z$ plots of our matrix elements (interpreted as continuous curves) alongside the physical limit.

One approach to extracting the pion DA is to assume a possible functional form for $f$ and fit our continuum-physical matrix elements to Equation \ref{eq:continuum_physical_me}. We require the DAs to vanish outside of the physical (momentum space) region $x=[0,1]$, and to-date, our most-used candidate function is the common meson PDF global-fitting form:
\begin{equation}
    f_{m,n} = x^m (1-x)^n / B(m+1,n+1)
    \label{eq:functional_fitting_form}
\end{equation}
where $m$ and $n$ are undetermined constants that we obtain through fitting, and we divide by the beta function to normalize the lightcone DA.

\begin{figure*}
    \center
    \includegraphics[width=0.49\textwidth]{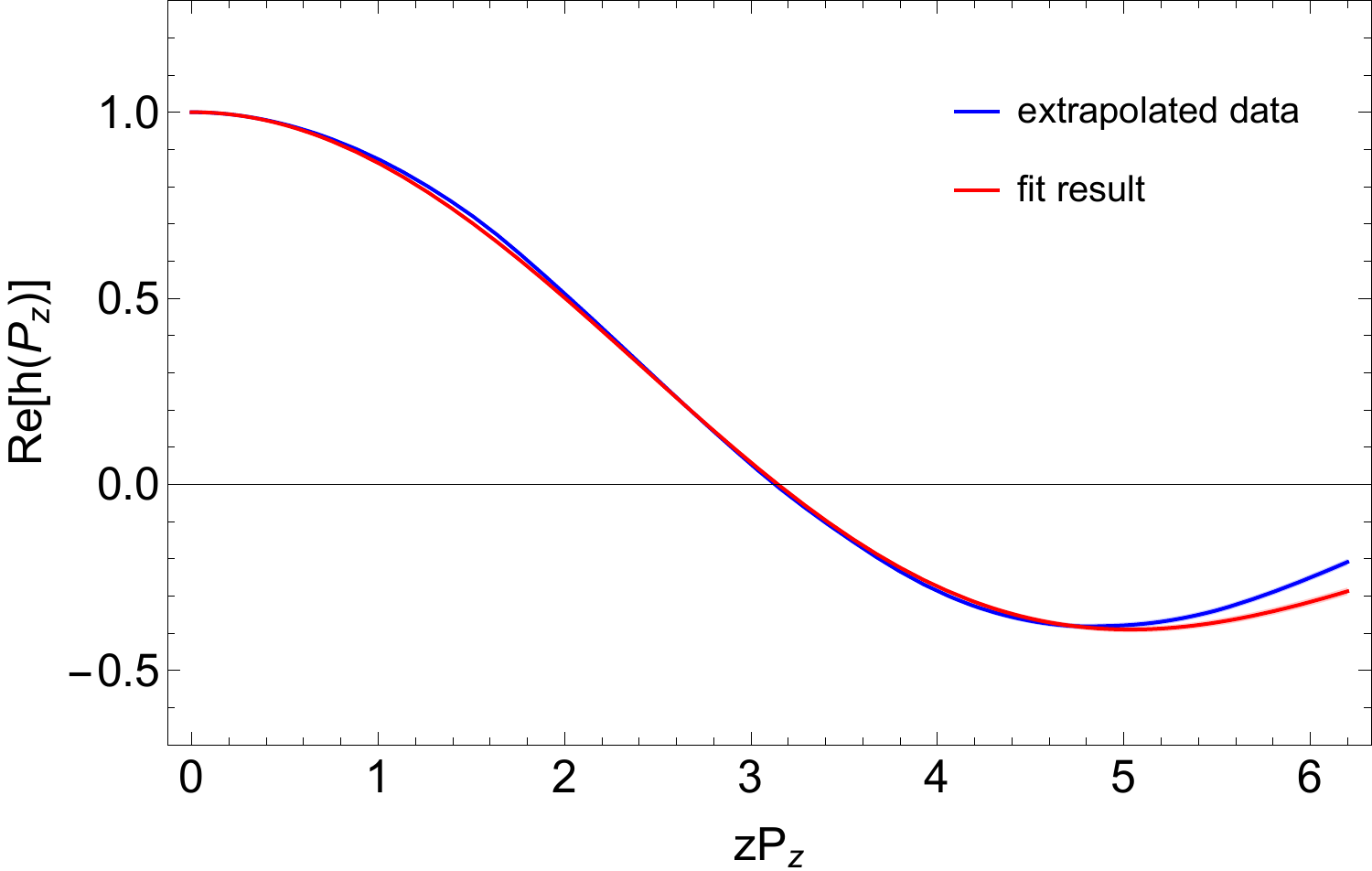}
    \includegraphics[width=0.49\textwidth]{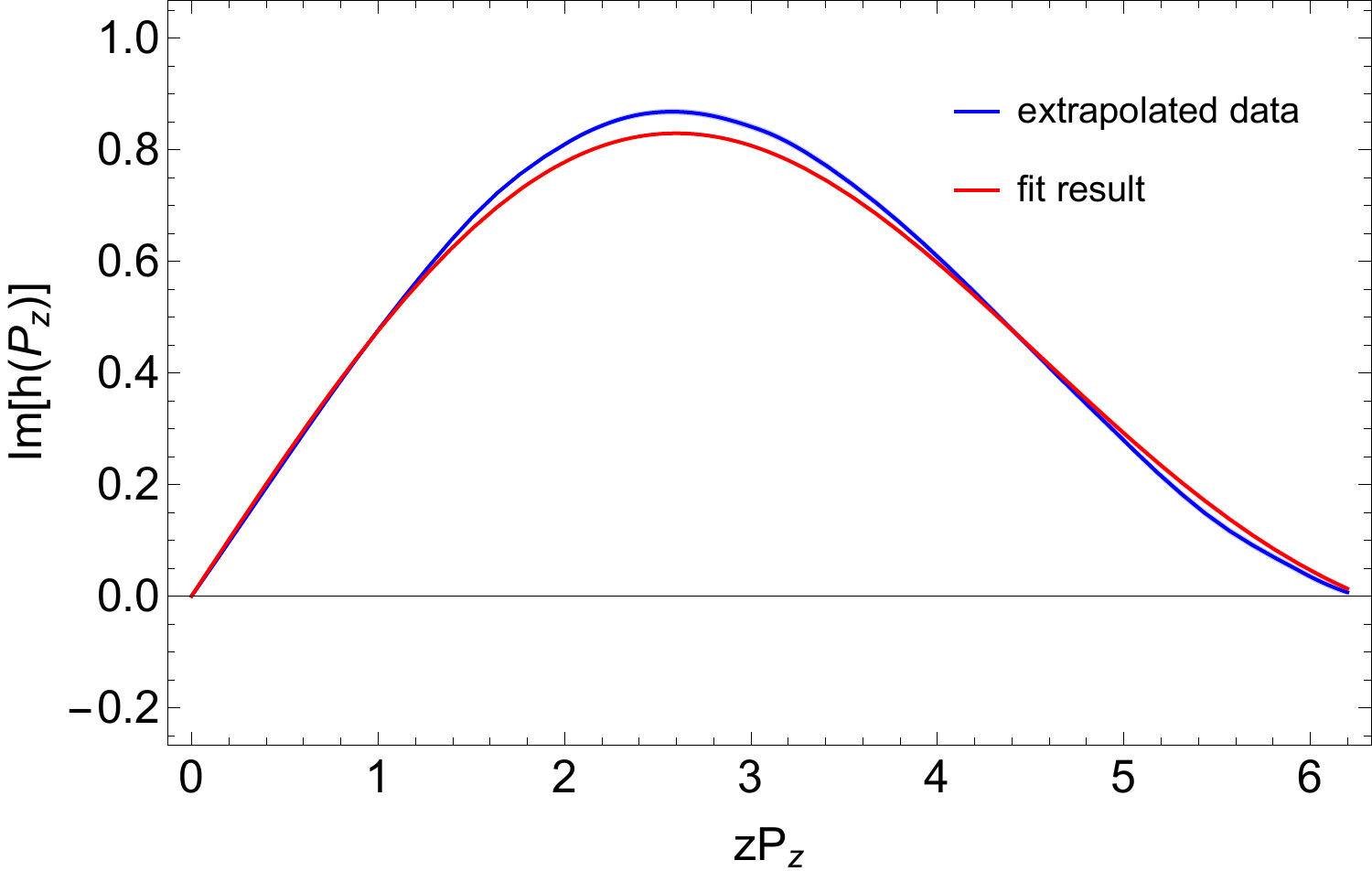}
    \includegraphics[width=0.49\textwidth]{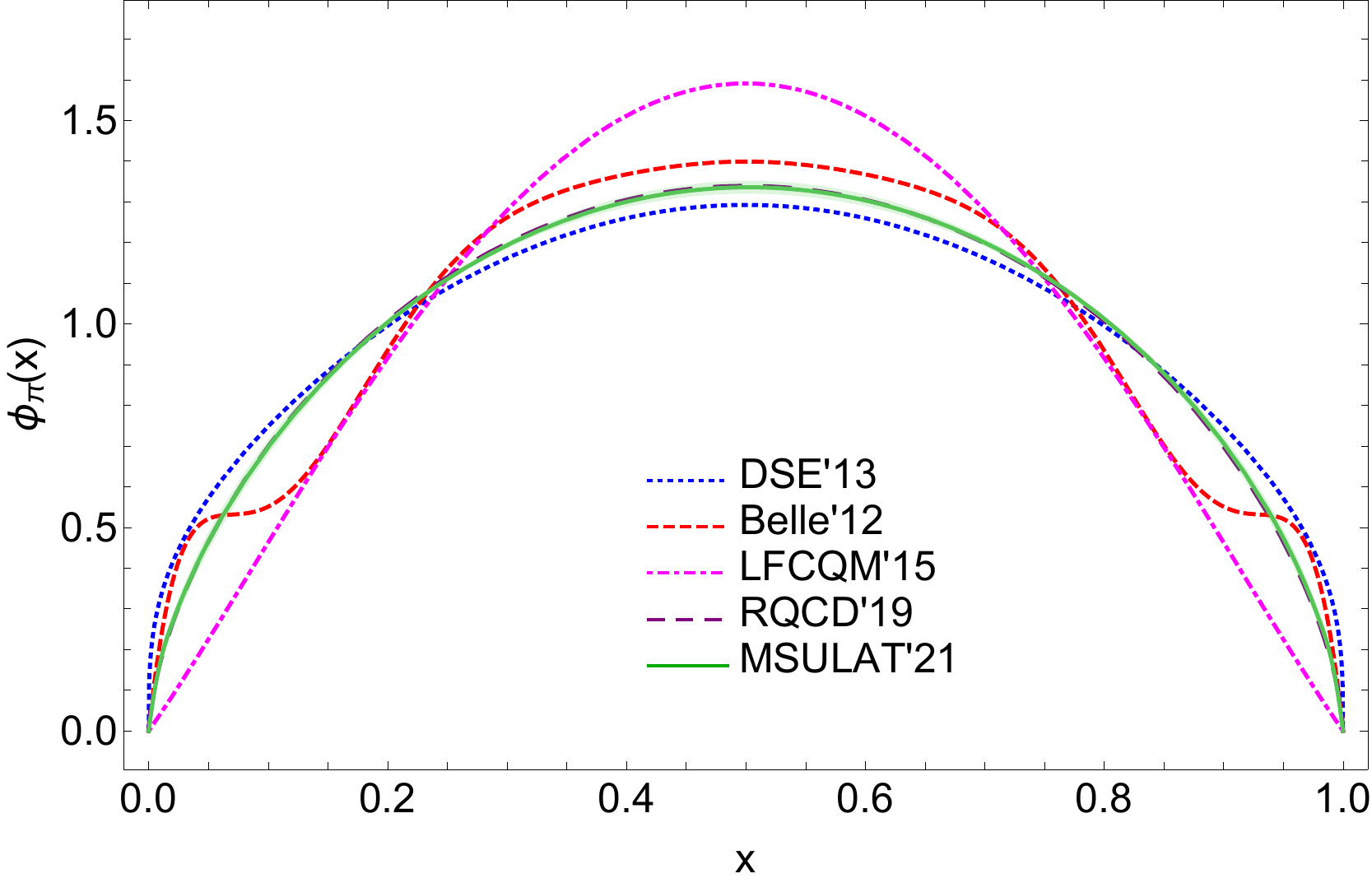}
    \caption{Results of the functional fit to the form in Equation \ref{eq:functional_fitting_form}. The top plots show fit result in position space against the extrapolated data. In the bottom plot, our preliminary pion lightcone DA is plotted in green against the results of earlier calculations.
    \label{fig:functional_fit}}
\end{figure*}

Figure \ref{fig:functional_fit} shows the corresponding fit results in position space using the extrapolated data at a momentum of 1.74 GeV, with the extrapolated data shown in blue. The real plot indicates a reasonable fit, but the imaginary plot shows a slight mismatch in the mid-$z$ region. Figure \ref{fig:functional_fit} also shows our preliminary pion lightcone DA in momentum space. We have plotted the results of previous calculations for comparison. We note that our lightcone DA shows a strong correlation with the $RQCD$'19 calculation.

Alternatively, we continue to \cite{Zhang:2020} use a machine learning (ML) approach to obtain the x-dependence of the pion DA. Since we cannot solve for $f_{m,n}$ directly from Equation \ref{eq:continuum_physical_me}, we can try to predict the form of the lightcone DA with a trained ML model. We train a multilayer perceptron (MLP) regressor with a set of pseudo-random polynomials $f_{m,n}$. Each function is of the same PDF global-fitting form used in the functional fitting approach with randomly generated $m,n$ pairs. 

We try two different regressor models. First, we use the standard MLP regressor as it is implemented in the scikit-learn Python package. We also use a modified version of this regressor with a custom loss function $L$:
\begin{equation}
    L = \frac{1}{2} \left[ \left( \frac{(y - y_\text{pred})}{y^{\gamma}} \right)^2 + \beta \left( \dfrac{d^2 y}{dx^2} \right)^2 \right]
    \label{eq:ML_custom_loss_func}
\end{equation}
where $\gamma$ and $\beta$ are tunable hyperparameters (this loss function is a generalization of the default mean-squared-error loss function). All of our regressors have 3 hidden layers with 20 nodes each. 

\begin{figure*}
    \center
    \includegraphics[width=0.49\textwidth]{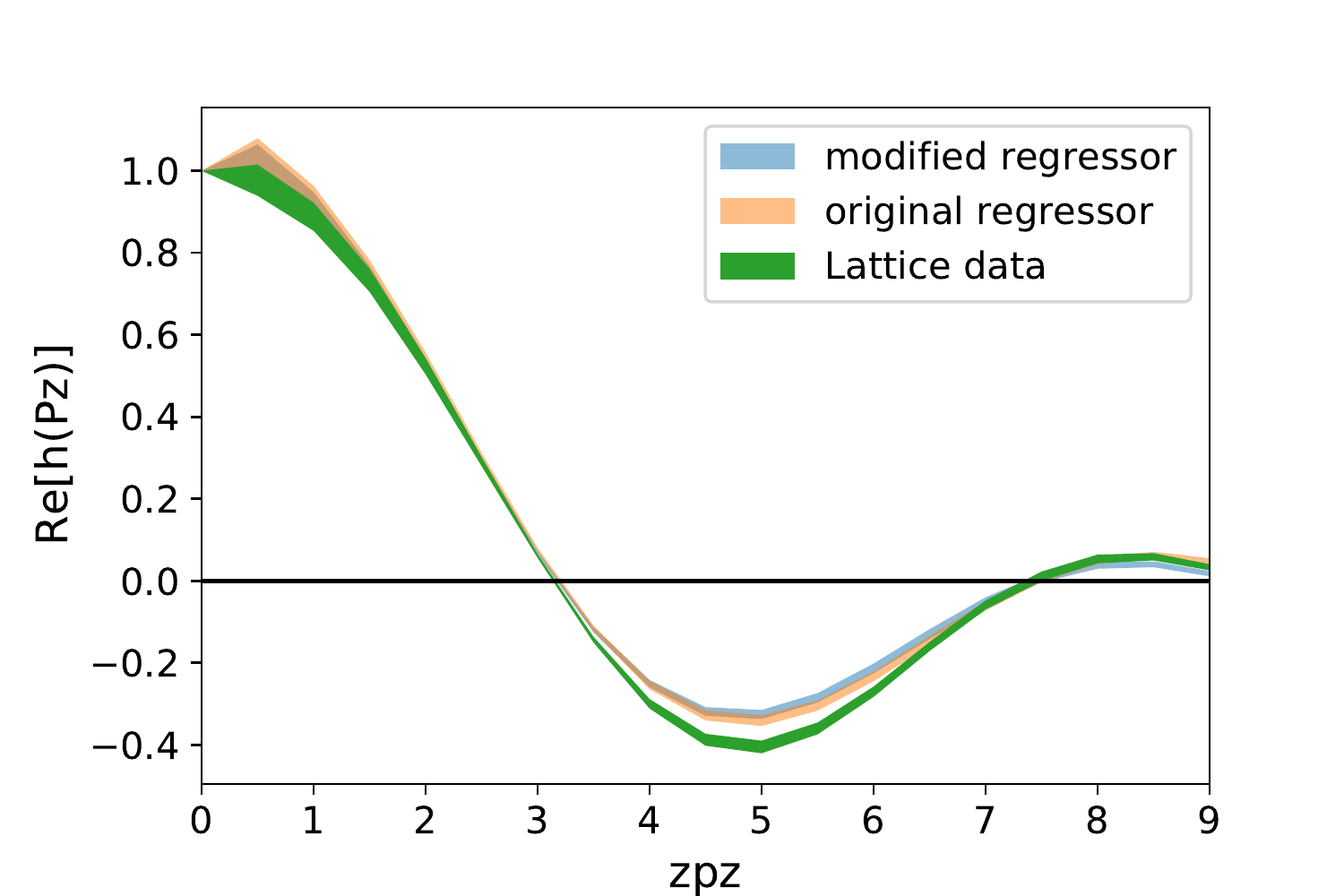}
    \includegraphics[width=0.49\textwidth]{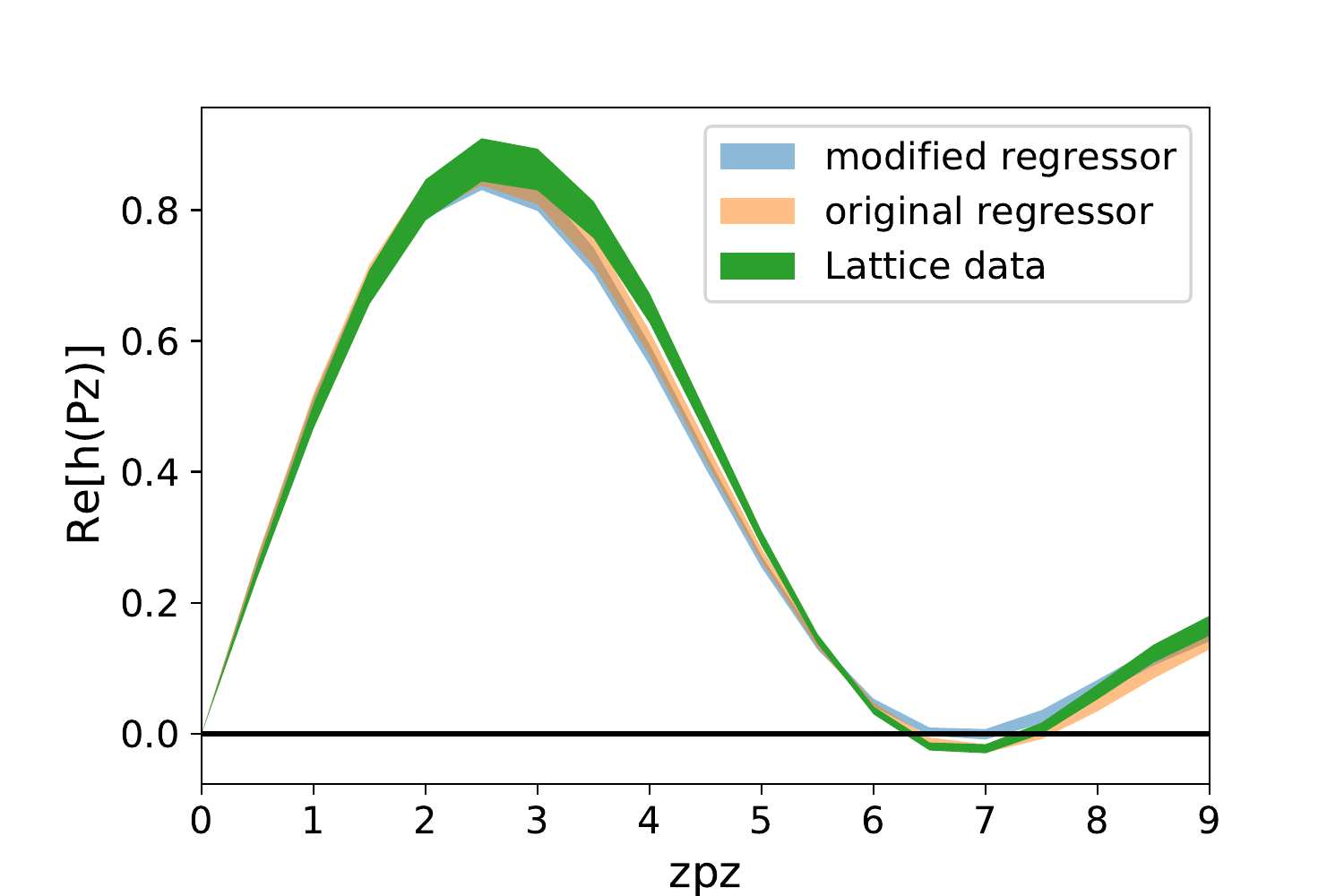}
    \includegraphics[width=0.49\textwidth]{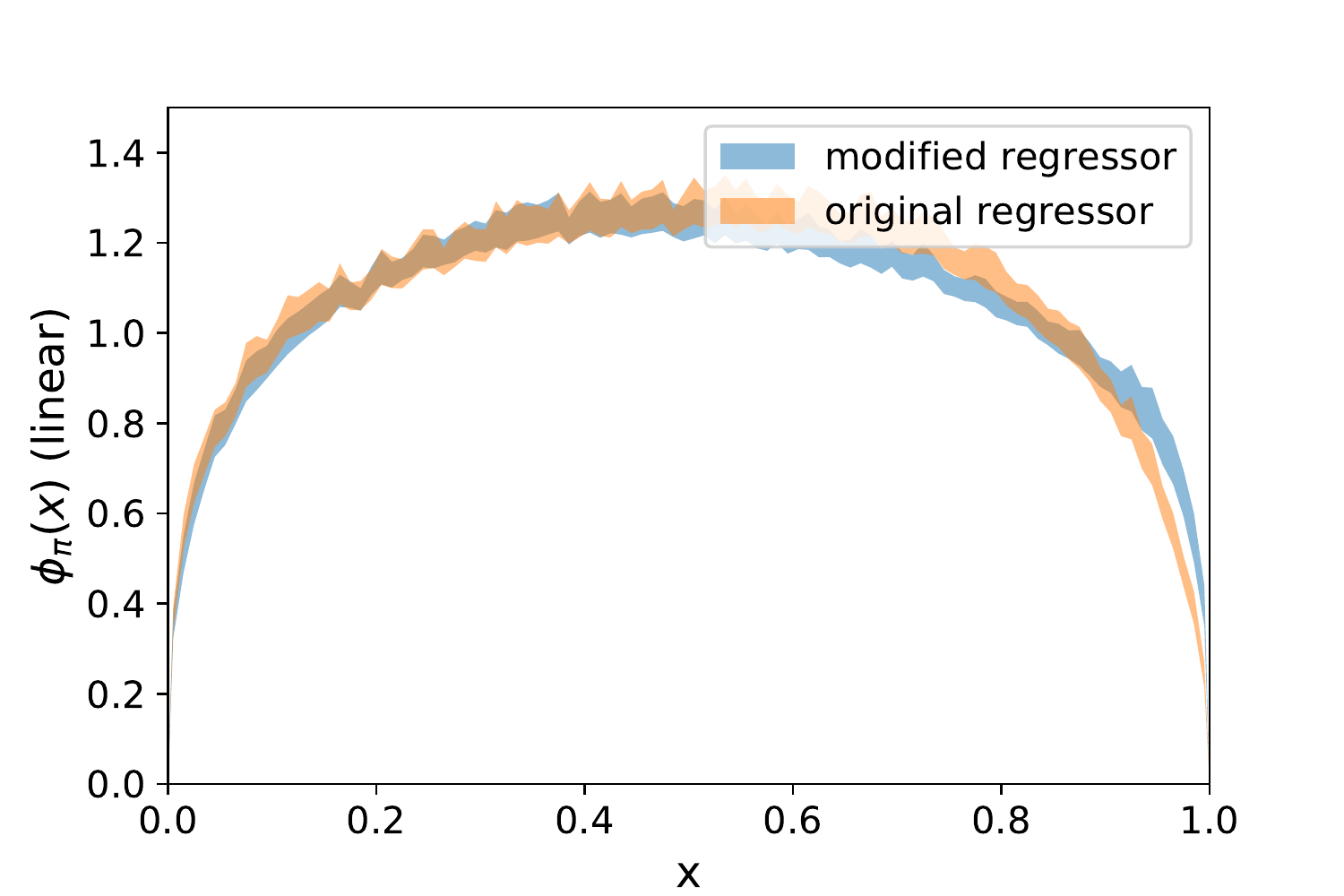}
    \caption{Machine learning predictions. The $zP_z$ plots show our predicted matrix elements from both the unmodified (original) and modified models against the true lattice data. The two predictions are very similar, and they show reasonable agreement with the lattice data. The $x$-space plot shows our (preliminary) predicted pion lightcone DAs from each model. Again, both predictions are close, though there is some disagreement at the high end of the $x$ range.
    \label{fig:ML_pred}}
\end{figure*}

Figure \ref{fig:ML_pred} shows an example of our predictions on each type of model. The two $zP_z$ plots show that our real and imaginary predicted MEs are fairly consistent with the true lattice data (with slight disagreement in the mid-$z$ region), while the two models predict only marginal differences from each other. The pion DA predictions resemble the functional fitting result, though the ML DA distributions are characteristically broader and flatter-- they are also noisier, which matches our expectations. Again, the two different models show agreement though they deviate a bit at larger $x$. This work is ongoing, and we still need to work out some of the kinks of the ML approach, partly with the treatment of the loss function.

\section*{Acknowledgments}

We thank MILC Collaboration for sharing the lattices used to perform this study. The LQCD calculations were performed using the Chroma software suite~\cite{Edwards:2004sx} with multigrid solvers~\cite{Babich:2010qb,Osborn:2010mb}. 
This research used resources of the
National Energy Research Scientific Computing Center, a DOE Office of Science User Facility supported by the Office of Science of the U.S. Department of Energy under Contract No. DE-AC02-05CH11231 through ERCAP;
facilities of the USQCD Collaboration, which are funded by the Office of Science of the U.S. Department of Energy,
and supported in part by Michigan State University through computational resources provided by the Institute for Cyber-Enabled Research (iCER).
The work of NJ is being supported by Graduate Fellowship of the college of  Natural Science at Michigan State University; CH is supported by the Professional Assistant program at Honors College at MSU. 
RZ and HL are partly supported by the US National Science Foundation under grant PHY 1653405 ``CAREER: Constraining Parton Distribution Functions for New-Physics Searches''.


\providecommand{\href}[2]{#2}\begingroup\raggedright\endgroup

\end{document}